\documentclass[11pt]{article}
\usepackage{amsmath}

\usepackage[normalem]{ulem}

\usepackage{float}
\usepackage{hhline}
\usepackage{color}
\usepackage{natbib}
\usepackage{booktabs}
\usepackage{threeparttable}
\usepackage{graphicx}
\usepackage{multirow}
\usepackage{amssymb}
\usepackage{lingmacros}
\usepackage{xspace}
\usepackage{pdflscape}
\usepackage{tabularx}
\newcolumntype{Y}{>{\centering\arraybackslash}X}

\newcommand{\be}{\begin{eqnarray}}
\newcommand{\ee}{\end{eqnarray}}
\newcommand{\ba}{\begin{eqnarray*}}
\newcommand{\ea}{\end{eqnarray*}}
\newtheorem{theorem0}{Theorem}
\newtheorem{lemma0}{Lemma}
\newtheorem{remark0}{Remark}
\newtheorem{fact0}{Fact}
\newtheorem{example0}{Example}
\newtheorem{corollary0}{Corollary}
\newtheorem{proposition0}{Proposition}
\newtheorem{conjecture0}{Conjecture}

\newenvironment{proposition}{\begin{proposition0}}{\end{proposition0}}

\def\boldfacefake #1{%
    \hbox{%
        \mathsurround=0pt
        \hbox to 0.4pt{$#1$\hss}%
        \hbox to 0.4pt{$#1$\hss}%
        \hbox {$#1$}%
    }%
}

\newcommand{\expect}{\mbox{\rm I\kern-.20em E}}
\newcommand{\reals}{\mbox{\rm I\kern-.20em R}}
\newcommand{\sreals}{\mbox{\small \rm I\kern-.20em R}}

\newcommand{\qed}{\nobreak \ifvmode \relax \else
      \ifdim\lastskip<1.5em \hskip-\lastskip
      \hskip1.5em plus0em minus0.5em \fi \nobreak
      \vrule height0.75em width0.5em depth0.25em\fi}

\begin{document}
\title{\bf Robust explicit estimation of the log-logistic distribution with applications}
\author{Zhuanzhuan Ma$^{a}$ \ \ and \ \ Min Wang$^{b,}$\footnote{Corresponding author. Email: min.wang3@utsa.edu} \ \ and \ \ Chanseok Park$^{c}$\\
{\small{$^{a}$ School of Mathematical and Statistical Sciences, The University of Texas Rio Grande Valley, Brownsville, TX}}\\
$^{b}${\small{Department of Management Science and Statistics, The University of Texas at San Antonio, TX}}\\
$^{c}${\small{Department of Industrial Engineering, Pusan National University, Busan, Korea}}
}

\date{}        
\maketitle

\begin{abstract}
The parameters of the log-logistic distribution are generally estimated based on classical methods such as maximum likelihood estimation, whereas these methods usually result in severe biased estimates when the data contain outliers. In this paper, we consider several alternative estimators, which not only have closed-form expressions, but also are quite robust to a certain level of data contamination. We investigate the robustness property of each estimator in terms of the breakdown point. The finite sample performance and effectiveness of these estimators are evaluated through Monte Carlo simulations and a real-data application. Numerical results demonstrate that the proposed estimators perform favorably
in a manner that they are comparable with the maximum likelihood estimator for the data without contamination and that they provide superior performance in the presence of data contamination.


\end{abstract}

\section{Introduction} \label{introduction}

The log-logistic (LL) distribution, also called the Fisk distribution in economics, is related to the logistic distribution in an identical fashion to how the log-normal and normal distributions are related to each other. {The LL distribution can be generated by taking an exponential transformation on the logistic distribution.} A random variable $T$ is said to follow the LL distribution with the scale parameter $\alpha > 0$ and the shape parameter $\beta >0$, denoted by  $T \sim \mathrm{LL}(\alpha, ~\beta)$, if its cumulative distribution function (cdf) can be written as
\begin{equation} \label{cdf:LL}
F(t \mid \alpha, \beta) = \frac{t^\beta}{\alpha^\beta + t^\beta},
\end{equation}
where $t > 0$, and its probability density function (pdf) takes the form
$$
f(t \mid \alpha, \beta) = \frac{\beta \alpha^{\beta} t^{\beta -1}}{(t^\beta + \alpha^\beta)^2}.
$$
It is worth noting that the LL distribution can be viewed as one of the commonly used parametric distributions for survival analysis. Unlike the well-known Weibull distribution, whose hazard rate function is $h(x \mid k, \lambda) = k\lambda^{-k}x^{k-1}$ with scale parameter $\lambda > 0$ and shape parameter $k > 0$, the LL distribution has a non-monotonic hazard rate function given by
\begin{equation*} \label{hazard:LL}
h(t \mid \alpha, \beta) = \frac{(\beta/\alpha)(t/\alpha)^{\beta -1}}{1 + (t/\alpha)^{\beta}},
\end{equation*}
which can increase initially and decrease later and at times can be hump-shaped. In addition, by allowing $\alpha$ to differ among groups, it can be adopted as the basis of an accelerated failure time (AFT) model. These statistical properties make the LL distribution be a favorable distribution in a variety of fields, such as economics (\citeauthor{fisk1961graduation} \citeyear{fisk1961graduation}), hydrology (\citeauthor{Shou:Mian:1988} \citeyear{Shou:Mian:1988}), medical and biological sciences (\citeauthor{geskus2001methods} \citeyear{geskus2001methods}), and engineering (\citeauthor{ashkar2003comparison} \citeyear{ashkar2003comparison}).

The parameters of the LL distribution are generally estimated based on classical methods such as maximum likelihood and least squares estimation methods. For instance, \cite{Bala:1987} studied the method of moments for the parameters of the truncated LL distribution. \cite{Kant:Srin:2002} studied the modified maximum likelihood estimation of $\alpha$ while assuming a known value of $\beta$. \cite{Reat:Wang:2018} proposed unbiased or nearly unbiased estimators for the parameters of the LL distribution. \cite{He:Chen:2020} reconsidered the maximum likelihood estimators based on simple random sampling and ranked set sampling techniques. More recently, \cite{He:2020} studied modified best linear unbiased estimator for estimating $\beta$. These existing estimators perform well for a complete data without any contamination, whereas they may become unreliable and can result in  severely distorted estimates when the data contain outliers. A small amount of data contamination even a single outlying observation
could induce a large impact on these estimators and even make them breakdown.

However, in many practical applications, there is no guarantee that the collected data are free of any contamination due to various volatile operating conditions when collecting data from an operating system. These observations motivate us to consider several outlier-resistant estimators for the parameters of the LL distribution, which include the percentile estimators, the repeated median estimators, the sample median and median absolute deviation estimators, the Hodges-Lehmann and Shamos estimators. It is worth noting that these estimators not only have simple closed-form expressions, but also are quite robust to a certain level of data contamination.
Furthermore, we investigate the breakdown point (\citeauthor{hampel1986robust} \citeyear{hampel1986robust}) of each estimator to measure its robustness in the presence of data contamination. Here, the breakdown point is defined as the proportion of incorrect observations (i.e. either arbitrarily large or small observations), the estimator of a parameter can deal with before yielding estimated values arbitrarily close to zero (implosion) or infinity (explosion). For example,  the mean has a breakdown point of $0\%$ and the median has a breakdown point of $50\%$.

The remainder of this paper is organized as follows. In Section \ref{section2}, we review maximum likelihood estimation (MLE) for the parameters of the LL distribution and provide a list of alternative outlier-resistant estimators. In Section \ref{section3},
we conduct extensive Monte Carlo simulations to investigate the finite sample performance of each estimator under consideration with and without data contamination. In Section \ref{section4}, we apply the proposed estimators to a real data example for illustrative purposes. Finally, some concluding remarks are provided in Section \ref{section5} with a proof deferred to the appendix.

\section{Parameter estimation methods} \label{section2}

Let $t_1, t_2, \cdots, t_n$ be $n$ observations from the LL distribution with the parameters $\alpha$ and $\beta$ in (\ref{cdf:LL}). The log-likelihood function can be written as
$$
\log L  = n\log\left(\beta\right) - n\log \left(\alpha \right) + \left(\beta-1\right)\sum_{i=1}^{n} \log t_i - 2 \sum_{i=1}^{n} \log \left[1 + \left(\frac{t_i}{\alpha}\right)^\beta \right].
$$
By taking the first derivative of the above log-likelihood function with respect to $\alpha$ and $\beta$, respectively, and then equating them to zero, the MLEs of $\alpha$ and $\beta$, denoted by $\hat \alpha$ and $\hat \beta$, can be obtained by finding the solutions of the following two equations
\begin{align*}
&2\sum_{i=1}^{n} \frac{ \left(t_i / \alpha\right)^\beta}{1 + \left(t_i / \alpha\right)^\beta} - n = 0,\\
&2\beta \sum_{i=1}^{n} \frac{\left(t_i / \alpha\right)^\beta \log\left(t_i / \alpha \right)}{1 + \left(t_i / \alpha\right)^\beta} - \beta \sum_{i=1}^{n} \log \left(t_i / \alpha  \right) - n = 0.
\end{align*}
Given that no explicit solutions are available to the two equations above, we can numerically obtain the MLEs $\hat \alpha$ and $\hat \beta$ by using the \texttt{mledist($\cdot$)} function from the {R} fitdistrplus package (\citeauthor{Marie:2015} \citeyear{Marie:2015}).

As mentioned above, the MLEs work well for estimating the parameters of the LL distribution when there is no data contamination, whereas they may result in unreliable and severely distorted parameter estimates in the presence of data contamination (even a single outlying observation); {see, for eample, \cite{park2003development, Park/Wang/Hwang:2020, Huber/Ronchetti:2009}}, among others. {This observation motivates us to develop several alternative  outlier-resistant estimators, including percentile estimators, repeated median estimators, sample median and median absolute deviation estimators, Hodges-Lehmann and Shamos estimators for the parameters of the LL distribution.}

\subsection{Percentile estimators}

By taking the inversion of LL cdf in (\ref{cdf:LL}), we obtain
\begin{equation} \label{cdf_inv}
t = \alpha\left[\frac{F(t\mid \alpha, \beta)}{1-F(t\mid \alpha, \beta)}\right]^{1/\beta}.
\end{equation}
To derive the percentile estimators, we may use the first ($t_{0.25}$) and third ($t_{0.75}$) quartiles of $n$ observations to solve for $\alpha$ and $\beta$ in (\ref{cdf_inv}) and obtain the following two equations
\begin{equation*}
t_{0.25} = \alpha \left(\dfrac{0.25}{1-0.25}\right)^{1/\beta} \  \mathrm{and} \ t_{0.75} = \alpha \left(\dfrac{0.75}{1-0.75}\right)^{1/\beta},
\end{equation*}
which provide the percentile estimators of $\alpha$ and $\beta$ given by
\begin{equation}
\hat{\alpha}_p= t_{0.25} \left(\dfrac{0.25}{1-0.25}\right)^{-1/\hat{\beta}_p} \ \mathrm{and} \
\hat{\beta}_p = \dfrac{2 \left[ \log \left(0.75\right) - \log\left(0.25\right)\right]}{\log \left(  t_{0.75} / t_{0.25} \right)},
\end{equation}
respectively. For any given pair of $n$ observations: low ($100l^{th}$) and high ($100h^{th}$), we can obtain the percentile estimators of $\alpha$ and $\beta$ given by
\begin{eqnarray} \label{ref:pe}
\hat{\alpha}_p &= t_{l} \left(\dfrac{l}{1-l}\right)^{-{1}/{\hat{\beta}_p}} \ \mathrm{and} \
\hat{\beta}_p & = \dfrac{2 \left[ \log \left(h\right) - \log\left(l\right)\right]}{\log \left(  t_{h} / t_{l} \right)} ,
\end{eqnarray}
respectively. To study the robustness property of the percentile estimators in (\ref{ref:pe}) based on contaminated data, we assume that all observations from the LL distribution are distinct, which however happens with probability one, since the LL distribution is continuous. We obtain the breakdown points of the estimators in (\ref{ref:pe}) summarized in the following proposition with its proof provided in the appendix.
\begin{proposition} \label{prop:01}
The asymptotic breakdown point of the symmetrical percentile estimators $\hat\alpha_p$ of $\alpha$ is given by
$$
\kappa = \min(1- h, 2h-1).
$$
The highest breakdown point of $\hat\alpha_p$ is $1/3$ when $l=1/3$ and $h = 2/3$, and the corresponding highest breakdown point of the percentile estimator $\hat\beta_p$ of $\beta$ with $\hat\alpha_p$ is $2/3$.
\end{proposition}

\subsection{Repeated median estimators}

It is worth noting that we rewrite the cdf of the LL distribution in (\ref{cdf:LL}) as a linear regression form
\begin{equation} \label{linearform}
y_i = \beta_0 + \beta_1 x_i \quad \mathrm{for} ~i = 1, \cdots, n,
\end{equation}
where $y_i  = \log \left[\left(1 - F(t_{(i)} \mid \alpha, \beta) \right)^{-1}-1 \right]$,
$x_i = \log \left(t_{(i)} \right)$, $\beta_0  = -\beta \log \alpha$, $\beta_1  = \beta$, and
$t_{(1)}, \cdots, t_{(n)}$ are the order statistics (from the smallest to the largest) of $n$ observations from the LL distribution. To approximate $F(t_{(i)}  \mid \alpha, \beta)$ in $y_i$, we adopt the technique of \cite{Ross:2000} with $i/(n+1)$, which can be easily implemented by using the \texttt{ppoints($\cdot$)} function in {R} language (\citeauthor{R:2011} \citeyear{R:2011}). The least squares estimates of $\beta_1$ and $\beta_0$ in (\ref{linearform}) are given by
\begin{equation*}
\hat{\beta}_1  = \dfrac{\sum_{i=1}^{n} (x_i - \bar{x_i})(y_i - \bar{y_i})}{\sum_{i=1}^{n} (x_i - \bar{x_i})^2}  \ \mathrm{and} \
\hat{\beta}_0  = \bar{y} - \hat{\beta}_1\bar{x},
\end{equation*}
respectively.
Analogous to the MLEs, they are not robust to data contamination with the breakdown points of $0\%$.
For this reason, we may consider the RM estimators (\citeauthor{siegel1982robust} \citeyear{siegel1982robust})
for estimating $\beta_0$ and $\beta_1$ in (\ref{linearform}), which are given by
\begin{equation} \label{RM:beta}
\hat \beta_1 = \mathop{\mathrm{med}}_{1 \le i \le n} \mathop{\mathrm{med}}_{j \neq i} \frac{ y_i - y_j}{ x_i - x_j}
  \quad \mathrm{and} \quad
\hat \beta_0 = \mathop{\mathrm{med}}_{1 \le i \le n} \big(y_i - \hat \beta_1 x_i \big),
\end{equation}
where med$_i(x_i)$ is the sample median for the $n$ observations $x_1, \cdots, x_n$. Then we obtain the RM estimators of the parameters of the LL distribution given by
\begin{equation} \label{RM:LL}
\hat{\alpha}_{\mathrm{RM}} = \exp \left( -\dfrac{ \hat{\beta}_{0}}{ \hat{\beta}_{1}} \right) \quad \mathrm{and} \quad
\hat{\beta}_{\mathrm{RM}}  =  \hat{\beta}_{1},
\end{equation}
respectively. It deserves mentioning that the breakdown points of the RM estimator in (\ref{RM:LL}) are equal to $50\%$, since the breakdown points of $\hat \beta_1$ and $\hat \beta_0$ in (\ref{RM:beta}) are $50\%$.

\subsection{Sample median and median absolute deviation estimators}
As mentioned in the introduction, if $ Z = \log T $, then $Z$ is said to follow a logistic regression distribution with the location parameter $\mu = \log \alpha$ and the scale parameter $s = 1/\beta$.
Therefore, the estimation of the LL parameters can be viewed as an estimation problem of $\mu$ and $s$ of
the logistic distribution based on $n$ observations $\{z_i: ~ z_i = \log t_i,~ t_i>0, ~i=1,2,\ldots,n \}$.

Since the parameter $\mu$ is the median of the logistic distribution, the sample median (SM) appears to be a natural choice for estimating $\mu$,  denoted by $\hat \mu = \mathop{\mathrm{med}}_{1 \le i \le n}\{z_1, \cdots, z_n\}$. A simple robust estimator for the scale parameter $s$ is the median absolute deviation (MAD) estimator given by
\begin{equation} \label{EQ:hats}
\hat{s} = \dfrac{\underset{1 \leq i \leq n}{\mathrm{med}}\{|z_i - \tilde{z}| \} }{\Phi^{-1}(3/4)},
\end{equation}
where $\tilde{z} = \mathrm{med}\{z_1, \cdots, z_n\}$, and {$\Phi^{-1}(\cdot)$} is the inverse of the standard normal cdf. Here, $\Phi^{-1}(3/4)$ is needed to make the estimator Fisher-consistent (\citeauthor{fisher1922mathematical} \citeyear{fisher1922mathematical}) for the standard deviation under the normal distribution (\citeauthor{hampel1986robust} \citeyear{hampel1986robust}, \citeauthor{park2003development} \citeyear{park2003development}). It deserves mentioning that the MAD estimator in (\ref{EQ:hats}) can be easily obtained using the \texttt{mad($\cdot$)} function from R language (\citeauthor{bellio2005introduction} \citeyear{bellio2005introduction}).
Then the sample median and MAD estimators of the parameters of the LL distribution are given by
\begin{equation} \label{MAD:LL}
\hat{\alpha}_{\mathrm{SM}} = \exp \left( \hat \mu \right) \ \mathrm{and} \
\hat{\beta}_{\mathrm{MAD}}  = \frac{1}{\hat s},
\end{equation}
respectively. {\cite{Park/Kim/Wang:2022} showed that that the asymptotic breakdown points of the estimators in (\ref{MAD:LL}) are equal to $50\%$ }.

\subsection{Hodges-Lehmann and Shamos estimators}
Within a location framework, \cite{hodges1963estimates} showed that the Hodges-Lehmann (HL) estimator performs better than the sample median for an asymmetric distribution in terms of efficiency. We may thus employ the HL estimator for estimating $\mu$ and it is defined as the median of $n(n-1)/2$ pairwise averages of $n$ observations given by
\begin{equation}
\tilde\mu = \underset{i < j}{\mathrm{med} } \left\lbrace\dfrac{z_i + z_j}{2} \right\rbrace,
\end{equation}
which can be easily obtained with the \texttt{hl.loc($\cdot$)} function in the R ICSNP package (\citeauthor{icsnp:2018} \citeyear{icsnp:2018}). Note that the HL estimator has a reasonable breakdown point of $29.3\%$ (\citeauthor{hettmansperger2010robust} \citeyear{hettmansperger2010robust}). We then adopt the Fisher-consistent Shamos estimator (\citeauthor{shamos1976geometry} \citeyear{shamos1976geometry}, \citeauthor{levy2011large} \citeyear{levy2011large}) for estimating  the scale parameter $s$ and it is defined as
\begin{equation} \label{EQ:Shamos}
\tilde{s} = \dfrac{\underset{i < j }{\mathrm{med}}\{|z_i - z_j| \} }{\sqrt{2}\Phi^{-1}(3/4)}.
\end{equation}
{To compute $\tilde s$, we let $\mathbf{Z}$ be the column vector such that $\mathbf{Z} = [z_1, \cdots, z_n]^T$. Similar to the outer product, we define the outer difference as below
\[
\mathbf{Z} \ominus \mathbf{Z}
= \begin{bmatrix}
z_{1}-z_{1}   &  z_{1}-z_{2}   &  \cdots & z_{1}-z_{n}    \\
z_{2}-z_{1}   &  z_{2}-z_{2}   &  \cdots & z_{2}-z_{n}    \\
\vdots        &  \vdots        &  \ddots & \vdots           \\
z_{n}-z_{1}   &  z_{n}-z_{2}   &  \cdots & z_{n}-z_{n}    \\
\end{bmatrix},
\]
which can be easily obtained using R language function
such as \texttt{outer(Z, Z, "-")}.
After obtaining the above outer difference, we take only lower or upper triangle matrix
which is also obtained by using \texttt{lower.tri()} or \texttt{upper.tri()}.
Then we have a set of $\mathcal{Z} = \{ z_{i}-z_{j} : i < j \}$
and take the absolute value of each element so that we have $\mathcal{W} = \{ |z_{i}-z_{j}| : i < j \}$.
Then we take the median using all the elements in $\mathcal{W}$. We may also use the \texttt{shamos($\cdot$)} function in rQCC package to achieve the computation; see \cite{park2020package}.
}

It is worth noting that the Shamos estimator has the same breakdown point of $29.3\%$ as the HL estimator (\citeauthor{rousseeuw1993alternatives} \citeyear{rousseeuw1993alternatives}). Then the Hodges-Lehmann and Shamos estimators of the parameters of the LL distribution are given by
\begin{equation} \label{HL:Sh}
\hat{\alpha}_{\mathrm{HL}} = \exp \left( \tilde{\mu} \right) \ \mathrm{and} \
\hat{\beta}_{\mathrm{Shamos}}  = \frac{1}{\tilde s},
\end{equation}
respectively. It can be shown that the asymptotic breakdown points of the estimators in (\ref{HL:Sh}) are also equal to $29.3\%$. {We here refer the interested reader to \citep{Park/Kim/Wang:2022} for a detailed derivation}.

\section{Simulation study} \label{section3}

In this section, we carry out Monte Carlo simulations to investigate the finite sample performance of each considered estimator for the case without data contamination (Section \ref{subsection:01}) and with data contamination (Section \ref{subsection:02}). For the simplicity of notation, we let ``PE1'', ``PE2'' and ``PE3'' stand for the symmetrical percentile estimators with the $5$th and $95$th, the $10$th and $90$th,  the $33$th and $67$th, respectively.

\subsection{Numerical results without contamination} \label{subsection:01}
We generate random samples of size $n = 10, ~25, ~50, ~75, ~100$  for the LL distribution with  $\alpha = 1$ and $\beta = 1.5, ~2.5, ~5, ~10$ by using \texttt{rllogis($\cdot$)} from the {R} package actuar  (\citeauthor{dutang2008actuar} \citeyear{dutang2008actuar}). Without loss of generality, we may fix $\alpha$ at $1.0$, since it is just a scale parameter. We replicate each simulation setting $M = 10,000$ times and compute the average bias and root mean square error (RMSE) of each estimator given by
\begin{equation*}
\begin{aligned}
\mathrm{Bias} = \frac{1}{M}\sum_{i=1}^M\big(\hat w_i - w_T\big) \quad
& \mathrm{and} \quad \mathrm{RMSE} = \sqrt{\frac{1}{M}\sum_{i=1}^M\big(\hat w_i - w_T\big)^2},
\end{aligned}
\end{equation*}
where $\hat w_i$ and $w_T$ stand for the estimate and true parameter for $w_T = \{\alpha, \beta\}$, respectively.

\begin{table}[!htbp]
\renewcommand{\arraystretch}{0.9}
\centering 
\small\addtolength{\tabcolsep}{-0.3pt}
\caption{Average bias and RMSE (in parentheses) of estimates for the scale  parameter $\alpha$ based on 10,000 simulations}
\label{table:sim01}
\begin{tabularx}{0.9\textwidth}{cc *{7}{Y}}
\multicolumn{4}{r}{} \\ \hline \hline
          &           & \multicolumn{7}{c}{Estimation methods}       \\ \cline{3-9}
$n$       &$\beta$   &   MLE  & PE1  &   PE2 &   PE3  &  RM &  SM  &  HL   \\ \hline
\multirow{6}{*}{10}
&\multirow{2}{*}{1.5}&$0.074$ &$0.299$&$0.179$&$0.085$&$0.065$&$0.088$&$0.076$\\[-0.08in]
          &           & (\textbf{0.424}) &  (0.951)& (0.645)& (0.453)& (0.425)& (0.472)& (0.430)\\[-0.04in]
&\multirow{2}{*}{2.5} & $0.026$  & $0.099$  & $0.058$  & $0.030$ & $0.021$  & $0.031$ & $0.027$ \\[-0.08in]
          &           & (\textbf{0.233})     & (0.381) & (0.303) & (0.247) & (0.235)  &  (0.257) &  (0.237)  \\[-0.04in]
&\multirow{2}{*}{5.0} & $0.007$     & $0.024$ & $0.014$  & $0.008$ & $0.004$ & $0.008$  & $0.007$  \\[-0.08in]
          &           & (\textbf{0.113})     &  (0.160) &  (0.139) &  (0.119) &  (0.114)  & (0.123) & (0.114)   \\[-0.04in]
&\multirow{2}{*}{10.0} &  $0.002$    & $0.006$ & $0.004$ & $0.002$ & $0.000$  & $0.002$ & $0.002$  \\[-0.08in]
          &           & (\textbf{0.056})     & (0.077) &  (0.068) &  (0.059) & (0.057)  &  (0.061) &  (0.057)  \\[-0.01in] \hline
\multirow{6}{*}{25}
&\multirow{2}{*}{1.5}& $0.027$ & $0.097$ & $0.064$  & $0.030$ & $0.023$  & $0.035$ & $0.027$  \\[-0.08in]
          &           & (\textbf{0.241})     & (0.450) & (0.339) &  (0.256) & (0.242)  &  (0.280) & (\textbf{0.241}) \\[-0.04in]
&\multirow{2}{*}{2.5} & $0.009$  & $0.034$ & $0.023$& $0.010$  & $0.007$ & $0.013$  & $0.010$  \\[-0.08in]
          &           & (\textbf{0.140})    & (0.241) & (0.190) &  (0.149) & (0.141)  &  (0.162) & (0.141) \\[-0.04in]
&\multirow{2}{*}{5.0} & $0.002$   & $0.009$ & $0.006$ & $0.002$ & $0.001$ & $0.003$& $0.002$  \\[-0.08in]
          &           & (\textbf{0.069})    & (0.115) &  (0.092) & (0.073) &  (0.070)  & (0.080) & (0.070)  \\[-0.04in]
&\multirow{2}{*}{10.0} & $0.001$  & $0.002$ & $0.001$ & $0.001$ & $0.000$  & $0.001$ & $0.001$  \\[-0.08in]
          &           & (\textbf{0.035})    & (0.057) &  (0.046) &  (0.037) & (\textbf{0.035})  & (0.040) & (\textbf{0.035})  \\[-0.01in] \hline
\multirow{6}{*}{50}
&\multirow{2}{*}{1.5} & $0.014$ & $0.058$  & $0.029$ & $0.017$ & $0.013$ & $0.019$ & $0.015$  \\[-0.08in]
          &           & (\textbf{0.167})   & (0.314) & (0.236) &(0.178) & (0.168)  & (0.192) &  (0.168)  \\[-0.04in]
&\multirow{2}{*}{2.5} &  $0.006$  & $0.021$ & $0.011$  & $0.007$ & $0.004$  & $0.007$ & $0.006$  \\[-0.08in]
          &           & (\textbf{0.099})     &  (0.177) &  (0.137) &  (0.105) & (\textbf{0.099})  &(0.113) &  (\textbf{0.099})  \\[-0.04in]
&\multirow{2}{*}{5.0} & $0.002$    & $0.006$  & $0.003$ & $0.002$ & $0.001$ & $0.002$ & $0.002$ \\[-0.08in]
          &           &  (\textbf{0.049})    &  (0.086) & (0.068) &  (0.052) & (\textbf{0.049})  &  (0.056) &  (\textbf{0.049}) \\[-0.04in]
&\multirow{2}{*}{10.0} & $0.000$ & $0.002$  & $0.001$ & $0.001$ & $0.000$ & $0.001$  & $0.000$   \\[-0.08in]
          &           &  (\textbf{0.024})    & (0.043) &  (0.034) & (0.026) &  (0.025)  & (0.028) & (0.025)  \\[-0.01in] \hline
\multirow{6}{*}{75}
&\multirow{2}{*}{1.5}& $0.009$   & $0.038$  & $0.021$ & $0.011$ & $0.009$  & $0.013$  & $0.010$  \\[-0.08in]
          &           & (\textbf{0.136})     &  (0.253) &  (0.190) & (0.144) &  (0.137)  & (0.158) &  (\textbf{0.136}) \\[-0.04in]
&\multirow{2}{*}{2.5} & $0.004$  & $0.014$ & $0.008$ & $0.004$& $0.003$& $0.005$  & $0.004$ \\[-0.08in]
          &           & (\textbf{0.080})    &  (0.146) &  (0.112) &  (0.085) &  (0.081)  &  (0.093) & (0.081)   \\[-0.04in]
&\multirow{2}{*}{5.0} & $0.001$    & $0.004$ & $0.002$ & $0.001$ & $0.001$  & $0.001$ & $0.001$   \\[-0.08in]
          &           &  (\textbf{0.040})     &  (0.072) &  (0.055) &  (0.042) &  (\textbf{0.040})  &  (0.046) &  (\textbf{0.040}) \\[-0.04in]
&\multirow{2}{*}{10.0} & $0.000$   & $0.001$ & $0.001$ & $0.000$ & $0.000$  & $0.000$  & $0.000$ \\[-0.08in]
          &           &  (\textbf{0.020})     & (0.036) & (0.028) &  (0.021) &  (\textbf{0.020})  &  (0.023) &  (\textbf{0.020}) \\[-0.01in] \hline
\multirow{6}{*}{100}
&\multirow{2}{*}{1.5}& $0.007$  & $0.028$  & $0.015$ & $0.008$ & $0.006$  & $0.009$ & $0.006$  \\[-0.08in]
          &           & (\textbf{0.118})    &  (0.225) &  (0.167) & (0.124) &  (0.119)  &  (0.136) & (\textbf{0.118}) \\[-0.04in]
&\multirow{2}{*}{2.5} & $0.003$    & $0.011$ & $0.006$ & $0.003$ & $0.002$  & $0.003$ & $0.003$  \\[-0.08in]
          &           & (\textbf{0.070})    & (0.131) & (0.099) &  (0.074) &  (0.071)  & (0.081) & (\textbf{0.070})  \\[-0.04in]
&\multirow{2}{*}{5.0} & $0.001$  & $0.003$  & $0.002$ & $0.001$ & $0.000$  & $0.001$ & $0.001$   \\[-0.08in]
          &          &  (\textbf{0.035})     & (0.065) & (0.049) &  (0.037) & (\textbf{0.035}) &  (0.040) & (\textbf{0.035}) \\[-0.04in]
&\multirow{2}{*}{10.0} & $0.000$  & $0.001$ & $0.001$ & $0.000$ & $0.000$  & $0.000$ & $0.000$   \\[-0.08in]
          &           & (\textbf{0.017})    &  (0.032) &  (0.024) &  (0.018) &  (0.018)  &  (0.020) &  (\textbf{0.017})  \\ \hline \hline
\multicolumn{9}{l}{\footnotesize Note: Bold value indicates the smallest values of RMSE for all the considered estimators}
\end{tabularx}
\end{table}

\begin{table}[!htbp]
\renewcommand{\arraystretch}{0.9}
\centering 
\small\addtolength{\tabcolsep}{-0.3pt}
\caption{Average bias and RMSE (in parentheses) of estimates for the shape  parameter $\beta$  based on 10,000 simulations}
\label{table:sim02}
\begin{tabularx}{0.9\textwidth}{cc *{7}{Y}}
\multicolumn{4}{r}{} \\ \hline \hline
          &           & \multicolumn{7}{c}{Estimation methods}       \\ \cline{3-9}
$n$       &$\beta$   &   MLE  &    PE1  &   PE2  &   PE3  &  RM &  MAD  &  Shamos   \\ \hline
\multirow{6}{*}{10}
&\multirow{2}{*}{1.5}&$0.221$ &$0.570$&$0.450$&$0.743$&$-0.004$&$-0.334$&$-0.596$\\[-0.08in]
          &           & (0.578)& (0.874)& (0.808)& (1.836)& (\textbf{0.555})& (0.681)& (0.669)\\[-0.04in]
&\multirow{2}{*}{2.5} &$0.368$ & $0.936$ & $0.779$ & $1.245$ & $-0.006$ & $-0.556$ & $-0.993$ \\[-0.08in]
          &           &  (0.963)   &  (1.440) & (1.362) & (3.065) & (\textbf{0.926}) & (1.136) & (1.115) \\[-0.04in]
&\multirow{2}{*}{5.0} & $0.735$  & $1.857$& $1.589$ & $2.496$  & $-0.012$ & $-1.113$ & $-1.986$  \\[-0.08in]
          &           & (1.926) & (2.862) & (2.739) & (6.131) & (\textbf{1.851})&(2.272)& (2.231) \\[-0.04in]
&\multirow{2}{*}{10.0} & $1.470$  & $3.705$ & $3.196$ & $4.996$& $-0.025$ & $-2.226$ & $-3.972$ \\[-0.08in]
          &           & (3.852) &  (5.716) & (5.487) & (12.258)& (\textbf{3.702}) & (4.543) & (4.462) \\[-0.01in] \hline
\multirow{6}{*}{25}
&\multirow{2}{*}{1.5}& $0.082$    & $0.254$& $0.179$ & $0.231$ & $-0.015$  & $-0.497$  & $-0.614$   \\[-0.08in]
          &           & (\textbf{0.291}) & (0.440) & (0.382) & (0.640) & (0.295) &(0.567) & (0.636) \\[-0.04in]
&\multirow{2}{*}{2.5} & $0.136$ & $0.421$ & $0.296$ & $0.386$  & $-0.025$  & $-0.828$ & $-1.023$   \\[-0.08in]
          &           & (\textbf{0.485}) &(0.733)&(0.636)&(1.067)&(0.492) &(0.945)&(1.059)\\[-0.04in]
&\multirow{2}{*}{5.0} & $0.272$   & $0.839$  & $0.592$  & $0.772$ & $-0.050$ & $-1.656$  & $-2.046$  \\[-0.08in]
          &           & (\textbf{0.970})   & (1.465) & (1.271) & (2.134)& (0.984) & (1.891) & (2.119) \\[-0.04in]
&\multirow{2}{*}{10.0} & $0.544$ & $1.676$ & $1.182$ & $1.544$ & $-0.101$  & $-3.313$ & $-4.092$   \\[-0.08in]
          &           & (\textbf{1.941})  & (2.930) & (2.541) &(4.268) &(1.968)& (3.782) & (4.238) \\[-0.01in] \hline
\multirow{6}{*}{50}
&\multirow{2}{*}{1.5} & $0.040$& $0.123$  & $0.092$  & $0.109$  & $-0.009$ & $-0.542$& $-0.619$ \\[-0.08in]
          &           & (\textbf{0.191})  &(0.265) &(0.248) & (0.384) & (0.204)  & (0.568) & (0.629)\\[-0.04in]
&\multirow{2}{*}{2.5} & $0.066$ & $0.204$ & $0.154$& $0.181$ & $-0.016$ & $-0.903$  & $-1.031$ \\[-0.08in]
          &           & (\textbf{0.318}) & (0.441) & (0.413) & (0.640) & (0.339)& (0.946)& (1.048) \\[-0.04in]
&\multirow{2}{*}{5.0} & $0.132$  & $0.408$ & $0.308$ & $0.363$ & $-0.031$& $-1.805$ & $-2.063$  \\[-0.08in]
          &           & (\textbf{0.636})  & (0.882) & (0.826) & (1.281) & (0.679) &  (1.893) & (2.096) \\[-0.04in]
&\multirow{2}{*}{10.0} & $0.264$  & $0.815$& $0.616$ & $0.725$& $-0.063$  & $-3.611$ & $-4.125$  \\[-0.08in]
          &           & (\textbf{1.272}) & (1.765) & (1.653) & (2.561)  &  (1.357)  & (3.785)&(4.192)\\[-0.01in] \hline
\multirow{6}{*}{75}
&\multirow{2}{*}{1.5}& $0.027$  & $0.083$ & $0.062$ & $0.071$ & $-0.007$ & $-0.554$ & $-0.620$ \\[-0.08in]
          &           & (\textbf{0.151})  & (0.206) & (0.192) & (0.291) & (0.163) & (0.570) & (0.626)  \\[-0.04in]
&\multirow{2}{*}{2.5} & $0.045$ & $0.139$ & $0.103$  & $0.119$ & $-0.011$ & $-0.923$ & $-1.033$ \\[-0.08in]
          &           & (\textbf{0.252})  & (0.344) & (0.319) &(0.484) & (0.272) & (0.951) & (1.044)  \\[-0.04in]
&\multirow{2}{*}{5.0} & $0.090$  & $0.279$& $0.205$  & $0.238$ & $-0.022$  & $-1.847$& $-2.067$  \\[-0.08in]
          &           &(\textbf{0.503}) & (0.688) & (0.639) & (0.968)& (0.543) & (1.901) & (2.088) \\[-0.04in]
&\multirow{2}{*}{10.0} & $0.179$  & $0.5572$ & $0.410$ & $0.476$ & $-0.045$ & $-3.693$ & $-4.134$   \\[-0.08in]
          &           & (\textbf{1.007})  & (1.376) & (1.277) & (1.936)& (1.086)& (3.803) & (4.177)  \\[-0.01in] \hline
\multirow{6}{*}{100}
&\multirow{2}{*}{1.5}& $0.020$  & $0.065$ & $0.047$  & $0.053$  & $-0.005$  & $-0.561$ & $-0.621$  \\[-0.08in]
          &           & (\textbf{0.129}) & (0.176) & (0.164) &(0.246) & (0.141) & (0.573)& (0.626) \\[-0.04in]
&\multirow{2}{*}{2.5} & $0.034$  & $0.108$ & $0.079$ & $0.088$ & $-0.008$ & $-0.935$& $-1.035$   \\[-0.08in]
          &           & (\textbf{0.215})  & (0.294) & (0.273) &(0.410) & (0.235) & (0.955) &(1.043)  \\[-0.04in]
&\multirow{2}{*}{5.0} & $0.068$ & $0.217$ & $0.157$ & $0.176$ & $-0.016$ & $-1.871$ & $-2.070$ \\[-0.08in]
          &          &  (\textbf{0.430}) & (0.587) &(0.547) &(0.821) & (0.470)& (1.910) & (2.086)  \\[-0.04in]
&\multirow{2}{*}{10.0} & $0.136$  & $0.434$ & $0.314$ & $0.351$ & $-0.033$& $-3.741$ & $-4.140$   \\[-0.08in]
          &           & (\textbf{0.860})  &(1.174) & (1.093) & (1.641) & (0.939)  & (3.820) & (4.171) \\ \hline \hline
\multicolumn{9}{l}{\small Note: Bold value indicates the smallest values of RMSE for all the considered estimators}
\end{tabularx}
\end{table}

Numerical results are provided in Tables \ref{table:sim01} and  \ref{table:sim02}. Several findings for the case without data contamination can be summarized as follows. (1) We observe that in terms of RMSE, the MLE of $\alpha$ outperforms  other estimators most often and that in terms of bias, the RM estimator of $\alpha$ is superior than other estimators in most cases. (2) When the sample size is small (e.g., $n=10$), the RM estimator of $\beta$ provides better performance than other estimators in terms of bias and RMSE, and the performance of the MLE for $\beta$ becomes superior in terms of RMSE, as $n$ increases. (3) The considered percentile estimators PEs of $\alpha$ usually perform worse than other estimators in terms of bias and RMSE, whereas they perform well for estimating of $\beta$. (4)  As the sample size $n$ becomes larger, the bias and RMSE of all the considered estimators decrease significantly, and all the estimators behave similarly.

\subsection{Numerical results with contamination} \label{subsection:02}

To study the robustness of the proposed estimators, we first generate $n = 25$ LL random variables with $\alpha = 1.0$, $\beta = 10$ as the reference distribution (no contamination case) and then consider the case with data contamination under the following four scenarios: (i) $10\%$ replacement outliers from the LL distribution with $\alpha = 1.0$ and $\beta = 0.1$; (ii)  $10\%$ replacement outliers from the LL distribution with $\alpha = 4.0$ and $\beta = 10$; (iii) $10\%$ replacement outliers from a uniform distribution on the interval $(0, ~20)$, and (iv) $10\%$ extreme contaminated values from a point mass distribution at the value of $50$.

\begin{table}[!htbp]
\centering 
\small\addtolength{\tabcolsep}{-0.3pt}
\caption{Average bias and RMSE (in the parentheses) of the scale ($\alpha$) and shape ($\beta$) estimators with $n=25$ under the five different simulation settings based on 10,000 simulations}
\label{table:sim03}
\begin{tabular}{lcccccccc}    \hline \hline
          &  MLE   & PE1  &   PE2  &   PE3  &  RM   & SM/MAD & HL/Shamos  \\ \hline
\multicolumn{7}{l}{\textit{No contamination from LL$(\alpha = 1.0$, $\beta = 10)$}} \\[-0.04in]
$\alpha$  \multirow{2}{*}{}
          & $0.001$  & $0.002$ & $0.001$ & $0.001$ & $0.000$  & $0.001$ & $0.001$   \\[-0.08in]
          & (\textbf{0.035})    & (0.057) &  (0.046) &  (0.037) & (\textbf{0.035})  & (0.040) & (\textbf{0.035})  \\[-0.01in]
$\beta$   \multirow{2}{*}{}
		  & $0.544$&  $1.676$ & $1.182$ & $1.544$ & $-0.101$  & $-3.313$ & $-4.092$   \\[-0.08in]
          & (\textbf{1.941})&  (2.930) & (2.541) &(4.268) &(1.968)& (3.782) & (4.238) \\[-0.01in]
\multicolumn{7}{l}{\textit{$10\%$ contamination from LL$(\alpha = 1.0$, $\beta = 0.1)$}} \\ [-0.04in]
$\alpha$  \multirow{2}{*}{}
          & $ -0.168$  & $-0.788$ & $-0.309$ & $0.001$ & $-0.006$ & $0.001$ & $-0.005$ \\[-0.08in]
          &(0.170)  & (0.788) & (0.310) & (0.037) & (\textbf{0.035}) & (0.040) & (\textbf{0.035})  \\[-0.04in]
$\beta$   \multirow{2}{*}{}
		  & $-8.836$  & $ -8.377$ & $ -6.160$ & $ 1.544$ & $-1.602$ & $-3.316$ & $ -4.821 $ \\[-0.08in]
          &(8.836)  & (8.377) & (6.164) & (4.268) & (\textbf{2.365}) & (3.784) & (4.932)  \\[-0.04in]
\multicolumn{7}{l}{\textit{$10\%$ contamination from LL$(\alpha=4.0$, $\beta=10)$} }      \\ [-0.04in]
$\alpha$  \multirow{2}{*}{}
         & $0.101$ & $0.804$ & $0.535$ & $0.058$ & $0.005$ & $0.051$ & $0.073$ \\[-0.08in]
          &(0.108) & (0.805) & (0.536) & (0.071) & (\textbf{0.036}) & (0.067) & (0.083)  \\[-0.04in]
$\beta$   \multirow{2}{*}{}
		  & $-4.824$  & $-5.995$ & $-5.963$ & $0.685$ & $-1.536$ & $-3.668$ & $-4.682$  \\[-0.08in]
          &(4.834)  & (5.996) & (5.966) & (3.746) & (\textbf{2.330}) & (4.061) & (4.800)  \\[-0.04in]
\multicolumn{7}{l}{\textit{$10\%$ contamination from U$(0,20)$} }      \\ [-0.04in]
$\alpha$  \multirow{2}{*}{}
          & $0.142$  & $2.191$ & $1.548$ & $0.058 $ & $0.005$ & $0.051$ & $0.073$  \\[-0.08in]
          &(0.148)  & (2.192) & (1.549) & (0.071) & (\textbf{0.036}) & (0.067) & (0.083)  \\[-0.04in]
$\beta$   \multirow{2}{*}{}
		  & $-6.959$  & $-7.745$ & $-7.912 $ & $0.685 $ & $-1.602$ & $-3.668$ & $-4.774$ \\[-0.08in]
          &(6.959) &   (7.745) & (7.912) & (3.746) & (\textbf{2.365}) & (4.061) & (4.888)  \\[-0.04in]
\multicolumn{7}{l}{\textit{$10\%$ contamination from a point mass distribution at $50$} }  \\ [-0.04in]
$\alpha$  \multirow{2}{*}{}
          & $ 0.104$ & $5.208$ & $3.988 $ & $0.001$ & $0.005$ & $0.001$ & $0.006$ \\[-0.08in]
          &(0.111)  & (5.213) & (3.991) & (0.037) & (\textbf{0.036}) & (0.040) & (\textbf{0.036})  \\[-0.04in]
$\beta$   \multirow{2}{*}{}
		  & $-7.965$  & $-8.589$ & $-8.786$ & $ 1.544$ & $-0.527$ & $-3.316$ & $-4.678$  \\[-0.08in]
          &(7.965)  & (8.589) & (8.786) & (4.268) & (\textbf{2.081}) & (3.784) & (4.797)  \\ \hline  \hline
\multicolumn{9}{l}{\footnotesize Note: Bold value indicates the smallest values of RMSE for all the considered estimators} \label{Table:03}
\end{tabular}
\end{table}

We conduct $M= 10,000$ replications for each scenario above and calculate the average bias and RMSE of all the estimators, which are summarized in Table \ref{Table:03}. Some conclusions for the case with data contamination can be drawn as follows. (1) We observe that data contamination by outliers induces a large influence on the MLE in terms of larger values of bias and RMSE. (2) The PE3 is superior than PE1 and PE2, and this occurs because the observations on PE1 and PE2 are outliers based on our simulation setups above. (3) The SM and HL estimators seem performing well for estimating $\alpha$, whereas they are badly affected by outliers when estimating $\beta$ for all the considered cases. (4) The RM estimator provides the best performance among all the considered estimators in most cases in terms of bias and RMSE. Finally, it deserves mentioning that we have conducted more simulation studies with respect to other values of the parameters and different sample sizes and obtained similar conclusions, which are thus not presented here for simplicity.

As a result, we observe that when the data contamination is present, the MLEs of the parameters are adversely affected by outliers. On the other hand, the proposed estimators are quite robust to a certain level of data contamination in terms of bias and RMSE, even when the sample size is small. Among the proposed robust estimators, numerical results showed that the RM estimators consistently provide more reliable results than others in most cases under consideration. Therefore, we have a preference for the RM estimators as alternative estimators to the MLEs for the parameters of the LL distribution, especially when the data contain outliers.

\section{A real-data application} \label{section4}

\begin{figure}[!htbp]
\hspace*{-0.8cm}
\includegraphics[width=0.55\textwidth, angle =-90]{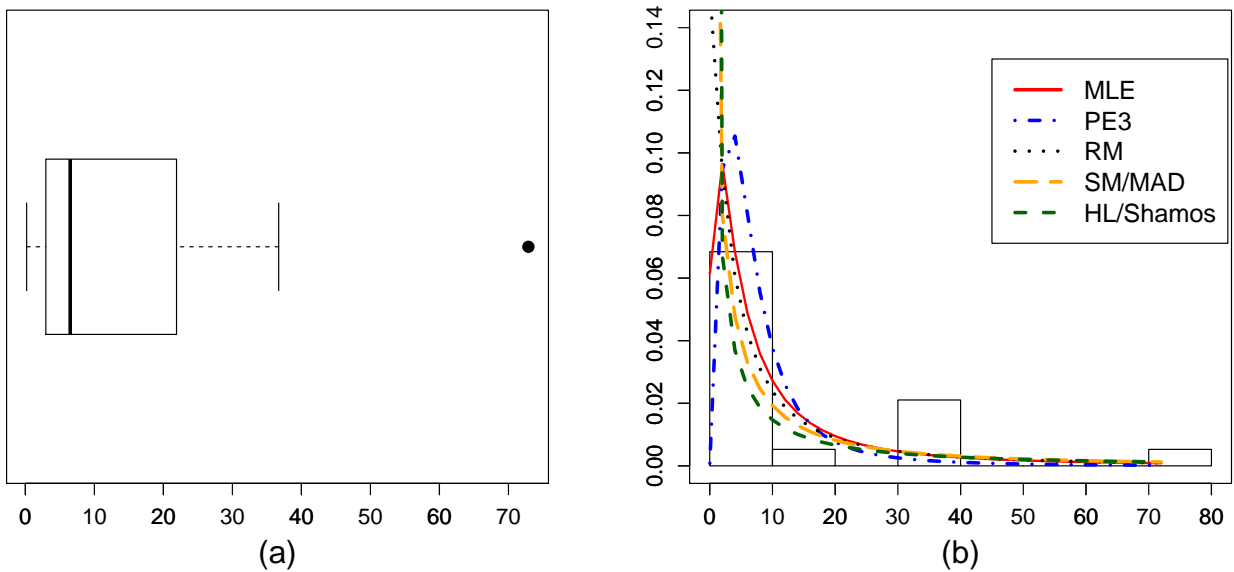}
\caption{(a) Box plot of the breakdown time data in \cite{nelson2003applied}; (b) Histogram with the fitted densities based on the considered estimators}
\label{fig1}
\end{figure}

In this section, we employ a real-data example to investigate the effectiveness of the proposed robust estimators. The data are originated from \cite{nelson2003applied} and are recently adopted by \cite{abbas2016objective} and \cite{Reat:Wang:2018} to illustrate the practical application of the LL distribution for this modeling lifetime data. The data record the breakdown time in minutes of an insulating fluid between electrodes at a voltage of $34$ kV. The data are provided in Table \ref{table:sim04}. It can be seen from the boxplot of the data in Figure \ref{fig1}(a) that there exists an outlier in this data.

\begin{table}[!htbp]
\centering 
\caption{The breakdown time in minutes of an insulating fluid data in \cite{nelson2003applied}}
\label{table:sim04}
\begin{tabular}{ccccccccccccccccc}   \hline  \hline
0.19 & 0.78 &  0.96&  1.31&  2.78& 3.16& 4.15 & 4.67& 4.85& 6.50 \\
7.35& 8.01& 8.27& 12.06& 31.75& 32.52& 33.91  & 36.71 & 72.89    \\ \hline  \hline
\end{tabular}
\end{table}

\begin{figure}[!htbp]
\begin{center}
\includegraphics[width=\textwidth]{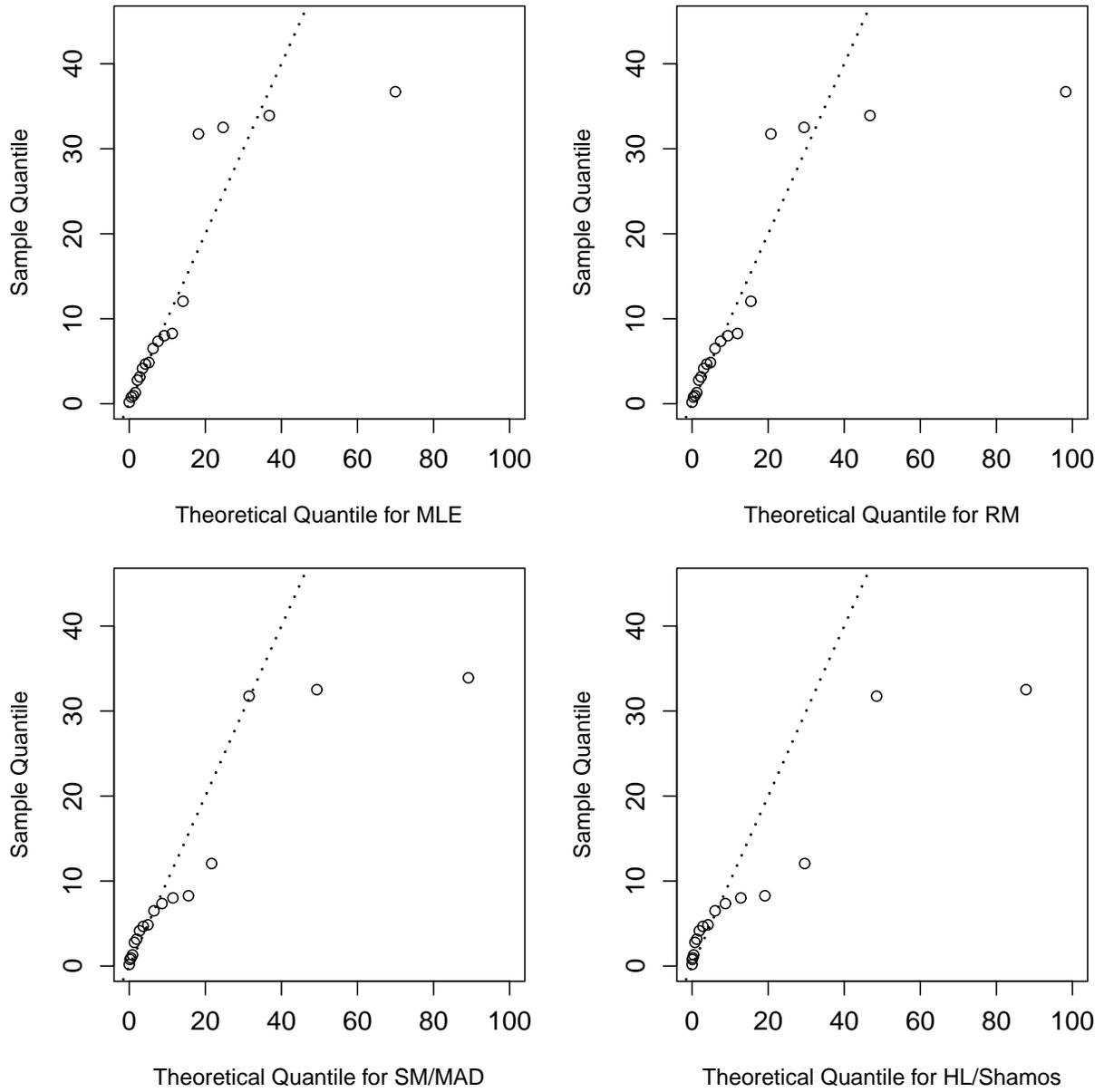}
\end{center}
\caption{The Q-Q plots for the breakdown time data (\citeauthor{nelson2003applied} \citeyear{nelson2003applied}) estimated by the considered estimation methods}
\label{fig2}
\end{figure}

The parameter estimates of $\alpha$ and $\beta$ and the Kolmogorov-Smirnov (KS) goodness of fit test (\citeauthor{smirnov1948table} \citeyear{smirnov1948table}) statistics are provided in Table \ref{table:sim05}. Here, the KS statistic is a nonparametric test that is often used to compare different estimation methods in terms of the quality of fit: the smaller the value
of this statistic, the better the fit to the data. We observe from Table \ref{table:sim05} that according to the KS test statistics, the best fit is obtained based on the RM estimators. Furthermore, we depict the histogram of the data with the fitted densities based on the MLE, PE3, RM, SM/MAD, and HL/Shamos estimation methods in Figure \ref{fig1}(b). We can visualize from this figure that the best fit of the data is also achieved by the RM method. Finally, the quantile-quantile (Q-Q) plots of the fitted LL distributions based on the considered estimators are presented in Figure \ref{fig2}. We observe that the quantile lines based on the MLE, SM/MAD, and HL/Shamos estimators show a tendency towards the outlier, whereas the RM estimator was not impacted. This real-data example further demonstrates the superior performance of the RM estimators over other considered estimators when the data contain outliers.

\begin{table}[!htbp]
\caption{Estimates of the parameters of the LL distribution for the breakdown time data}
\label{table:sim05}
\centering 
\begin{tabularx}{0.9\textwidth}{l *{3}{Y}}    \hline \hline
Estimation method    & $\hat{\alpha}$ & $\hat{\beta}$ &  KS ({$p$-value})  \\ \hline
MLE      & $6.2573$  & $1.1732$ &  $0.1337 ~({0.8428})$   \\
PE3      & $5.8957$  & $1.9374$ &  $0.2263 ~({0.2453})$       \\
RM       & $6.0318$  & $1.0153$ &  $\mathbf{0.1069} ~ ({0.9654})$  \\
SM/MAD   & $6.5000$  & $0.7941$ &  $0.1492 ~ ({0.7372})$ \\
HL/Shamos & $6.0429$ &$0.6014$  &  $0.1999 ~ ({0.3823})$  \\\hline \hline
\multicolumn{4}{l}{\footnotesize Note: The best value is marked in bold}
\end{tabularx}
\end{table}

\section{Concluding remarks} \label{section5}

In this paper, we have developed several alternative estimators to the MLEs for the parameters of the LL distribution, which not only have simple closed-form expressions, but also are quite robust to a certain level of data contamination by outliers. In addition, we have investigated the breakdown point of each estimator under consideration. Numerical results from simulation studies showed that in terms of bias and RMSE, the proposed estimators perform favorably in a manner that they are comparable to the MLEs in the absence of data contamination and that they provide superior performance in the presence of data contamination.

A natural question appears: which of the proposed robust estimators, the percentile estimator, the repeated median, the sample median and median absolute deviation estimators, or the Hodges-Lehmann and Shamos estimators should be recommended for estimating the parameters of the LL distribution when the data are contaminated by outliers. We have a preference for the RM estimators, because they consistently offer a better performance than other estimators under different simulation scenarios and also have a high breakdown point of $50\%$. Finally, a real-data application further illustrated that the RM estimators provide a better fit than other estimators in terms of the KS test statistic and the Q-Q plots.

\section*{Acknowledgments}
The authors would like to acknowledge the comments and suggestions
from two anonymous reviewers, which have improved the quality of the manuscript. This work was based on the first author's dissertation research which was supervised by the corresponding author. The work of Dr. Park was supported by the National Research Foundation of Korea (NRF) grant funded by the Korean government (NRF-2017R1A2B4004169).

\bibliographystyle{annals}

\section*{Appendix A} \label{appendixa}

\noindent{\textit{Proof of Proposition \ref{prop:01}:}}
It is worth noting that the estimator of the shape parameter  $\hat{\beta_p}$ in (\ref{ref:pe}) tends to infinity (explosion) when $t_h = t_l$. This could happen if a proportion ($h-l$) of the observations is placed to the same position as $t_l$. Also, the estimator $\hat{\beta_p}$ goes to zero (implosion) if $t_h$ approaches infinity and $t_l$ remains bounded or if $t_l$ goes to zero and $t_h$ remains bounded. In this case, it suffices to place ($1-h$) observations to go to infinity or $l$ observations to be zero. These observations indicate that the breakdown point of $\hat{\beta_p}$ is given by
\begin{align*}
\kappa = \min\{h, ~1-l, ~h-l, ~1-h\} = \min\{1-l, ~h-l, ~1-h\}.
\end{align*}
If we consider the symmetric percentiles (i.e. $h = 1-l$) and $h > 1/2$, the breakdown point of the symmetrical percentile estimator $\hat{\beta_p}$ is then given by
$$
\kappa = \min\{2h-1, ~1-h\}.
$$
The highest breakdown point for the symmetrical percentile estimator $\hat{\beta_p}$ is obtained at the intersection of lines $2h -1 = 1- h$ for $h = 2/3$ and $l = 1/3$. In a similar way as done for the shape estimator $\hat{\beta_p}$, we can show that the breakdown point of the scale estimator $\hat{\alpha_p}$ is equal to $\min\{h, 1-l \}$. Therefore, we obtain the highest asymptotic breakdown point the percentile estimator $\hat \alpha_p$ corresponding to the symmetrical percentile estimator $\hat \beta_p = 2/3$ when $h = 1-l= 2/3$. This completes the proof.\\

\section*{{Appendix B}}\label{appendixb}
\noindent{\textit{{{R} code for the real-data application in Section \ref{section4} :}}}

\begin{verbatim}
library("fitdistrplus") # mledist
library("rQCC")
set.seed(1)
# The time to breakdown of an insulating fluid
X <- c(0.19, 0.78, 0.96, 1.31, 2.78, 3.16, 4.15, 4.67, 4.85, 6.50,
       7.35, 8.01, 8.27, 12.06, 31.75, 32.52, 33.91, 36.71, 72.89)
n <- length(X)
x <- log(X)             # rewrite log-logistic to location-scale form
F <- ppoints(n, a=0)    # F = i/(n+1)
y <- log((1-F)^(-1)-1)  # rewrite log-logistic to linear regression model

# Maximum likelihood estimators
dLogL <- function(x, beta, alpha)(beta/alpha)*
         (x/alpha)^(beta-1)*((1+(x/alpha)^beta)^-2)  # log-logistic pdf
pLogL <- function(x, beta, alpha)
         (x^beta/(x^beta + alpha^beta))  # log-logistic cdf
fit.mle <- mledist(X, "LogL", start = list(alpha = 0.2, beta = 0.01))
fit.mle$estimate

# Percentile estimators
H <- 0.67
L <- 0.33
betahat.percentile <- function(data, H, L){
  PH <- quantile(data, H)
  PL <- quantile(data, L)
  betahat.percentile <- 2*(log(H) - log(L))*(log(PH/PL))^(-1)
  return(betahat.percentile)
}
betahat.pct <- betahat.percentile(X, H, L)
alphahat.pct<- exp(log(quantile(X, H)) - 1/betahat.pct * log(H/L))

# Repeated median estimators
beta1hat.med <- function(x, y) {
  int_list <- final_list <- numeric(0)
  for (j in 1:n) {
    for (i in 1:n) {
      if (i != j) {
        int_list = append(int_list, (y[j] - y[i]) / (x[j] - x[i]))
      }
    }
    final_list <- append(final_list, median(int_list))
  }
  return(median(final_list))
}
beta1hat.RM <- beta1hat.med(x, y)
beta0hat.RM <- median(y - beta1hat.RM*x)
betahat.rm <- mean(beta1hat.RM)
alphahat.rm<- exp(-beta0hat.RM/beta1hat.RM)

# Median absolute deviation (MAD) & Shamos estimators for alpha (scale)
# && sample median (SM) & Hodges-Lehmann (HL) estimators for beta (shape)
shape_est <- function(data){
  alpha.mad <- 1/mad(data)
  alpha.shamos <- 1/shamos(data)
  shape_est.result <- cbind(alpha.mad, alpha.shamos)
  return(shape_est.result)
}
betahat.mad <- shape_est(x)[,1]
betahat.shamos <- shape_est(x)[,2]

scale_est <- function(data){
  beta.median <- exp(median(data))
  beta.hl <- exp(HL(data, "HL2"))
  scale_est.result <- cbind(beta.median, beta.hl)
  return(scale_est.result)
}
alphahat.median <- scale_est(x)[,1]
alphahat.hl <- scale_est(x)[,2]
\end{verbatim}

\end{document}